\pdfoutput=1
\documentclass[a4paper,fleqn]{cas-sc}
\usepackage{amsmath}
\usepackage{amssymb}
\usepackage{comment}
\usepackage[numbers]{natbib}
\usepackage{lineno}

\begin{document}
\let\WriteBookmarks\relax
\def\floatpagepagefraction{1}
\def\textpagefraction{.001}

\shorttitle{A recurrent epidemic model with antigenic variation}
\shortauthors{R. Kumata et~al.}

\title[mode=title]{Immune history shapes recurrent epidemics of antigenically related variants}

\author[1,2,3]{Ryuichi Kumata}
\cormark[1]
\ead{ryuichi.kumata@cefe.cnrs.fr}
\orcidauthor{https://orcid.org/0000-0002-6428-6357}{R. Kumata}
\author[2]{Yuma Fujimoto}
\author[2]{Hisashi Ohtsuki}
\author[2]{Akira Sasaki}

\affiliation[1]{organization={CEFE, CNRS, Univ Montpellier, EPHE, IRD},
  country={France}}
\affiliation[2]{organization={Research Center for Integrative Evolutionary Science, SOKENDAI},
  country={Japan}}
\affiliation[3]{organization={Institute of Tropical Medicine, Nagasaki University},
  country={Japan}}

\cortext[1]{Corresponding author.}

\begin{abstract}
Population immunity carried over from past epidemics of an antigenically variable pathogen influences the epidemic of new variants based on their antigenic similarity to the previous ones. We develop a recurrent SIR model where a population faces sequential, antigenically related variants. The model yields a recurrence map for the population susceptibility to successive variants under the assumption of status-based population immunity. The model reveals that stable, equal-sized recurrent epidemics occur across broad parameter ranges, but can be destabilized when transmission is strong and antigenic escape is limited, leading to period-$2$ or more, or even more complex epidemic dynamics. Epidemic size is maximized at an intermediate basic reproduction number: higher transmissibility boosts immediate infection but also enhances cross-immunity, reducing future susceptibility of the population. Our results clarify how immune history shapes recurrent epidemics and why success in one wave does not ensure larger future epidemics.
\end{abstract}


\maketitle

%\linenumbers

\section{Introduction}

Understanding the interplay between population immunity and epidemiological dynamics is fundamentally important for the prediction and control of infectious diseases \cite{metcalf2015understanding,gandon2016forecasting,saad2022immuno,grenfell2004unifying}. In a single epidemic caused by a pathogen, transmission depletes the susceptible population due to immunization, and this depletion of susceptible hosts ultimately ends epidemics. This is one of the central insights of classical epidemiology theory \cite{kermack1927contribution}. Many pathogens, however, evolve their antigenicity to escape the existing population immunity shaped by past epidemics and thereby cause recurrent epidemic waves. In such dynamics, an epidemic not only reduces susceptibility to the currently circulating variant, but also alters the immune environment encountered by antigenically related variants that appear later. Thus, the history of past epidemics may influence future epidemiological dynamics through the population immunity that is shaped by the epidemiological history.

This possibility is particularly relevant for antigenically evolving pathogens such as influenza and SARS-CoV-2. Influenza variants circulate through antigenic space while remaining partially connected by cross-immunity, and immune histories accumulated through previous exposure shape responses to later variants \cite{smith2004mapping,cobey2017immune}. Similarly, recent works on SARS-CoV-2 have shown that population immunity can help predict the evolutionary success of newly emerging variants \cite{meijers2023population,raharinirina2025sars}. These observations suggest a feedback: epidemic waves reshape population immunity, and population immunity in turn affects the chance for later variants to spread and cause epidemics. Although this feedback is biologically plausible and empirically relevant, it remains unclear how the cumulative immune footprint left by prior epidemic waves affects the epidemiological dynamics of antigenically related pathogens.

Previous theoretical studies have shown that recurrent epidemic dynamics can be shaped by feedback among epidemic size, cross-immunity, antigenic drift, immune waning, and demographic turnover \cite{andreasen2003dynamics,boni2004influenza,roberts2019simple}. Nonetheless, one important aspect shaping this feedback remains underexplored. When protection against a future variant depends on antigenic similarity, immune effects generated by different past epidemic waves need not accumulate in the same way. An earlier wave provides stronger protection against a future variant when
the two variants are antigenically closer than when they are more distant. It therefore remains unclear how these cumulative, variant-specific immune effects shape the stability and size of later recurrent epidemics.

Here, we address this problem with a recurrent epidemic model in which a focal host population is exposed sequentially to antigenically related variants. 
We condition on the sequence of introduced variants and assume that each variant causes a single SIR epidemic wave, reducing susceptibility to itself and to later variants according to antigenic similarity. 
This yields a recurrence map that accumulates the variant-specific immune effects of previous waves. 
We show that the strength and shape of cross-immunity determine the stability of equal-sized recurrent epidemics and the emergence of multi-periodic dynamics. 
We further show that recurrent epidemic size is maximized at an intermediate basic reproduction number, reflecting a trade-off between infection within the current wave and cross-immunity imposed on later variants.

\section{Model}

Our goal is to characterize how immune history generated by successive epidemic waves shapes the epidemiological conditions encountered by later antigenically related variants in a focal host population. We represent this process as a wave-to-wave mapping: each epidemic wave begins in the susceptibility landscape left by its predecessors, and its size determines the immunity inherited by the next wave. This mapping allows us to recursively determine the susceptible density at the onset of each wave, the within-wave susceptibility retention factor, and the corresponding epidemic size.

We assume that antigenically related variants circulate and evolve outside the focal population and are introduced into it intermittently. Accordingly, we do not model their generation, migration, or establishment; instead, we take the sequence of introductions and their antigenic relationships as given. The model therefore isolates the local immunological feedback through which each epidemic reshapes the conditions faced by subsequent variants.

Variants are indexed by their order of introduction, $i=0,1,2,\ldots$. When variant $i$ causes an epidemic, we refer to the resulting outbreak as wave $i$. The within-wave time is denoted by $t$, with $t=0$ at variant introduction, and the total host population density is normalized to one. At the beginning of wave $j$, let $S_i^{(j)}(0)$ be the density of hosts susceptible to variant $i$. The superscript denotes the current wave, whereas the subscript denotes the variant against which susceptibility is measured. Thus, $S_j^{(j)}(0)$ is susceptibility to the current epidemic variant, while $S_i^{(j)}(0)$ for $i>j$ describes susceptibility to a future variant. Initially, the population is fully susceptible to all variants, $S_i^{(0)}(0)=1$.

\subsection{Epidemiological dynamics within a single epidemic wave}

\begin{figure}
    \centering
    \includegraphics[width=\linewidth]{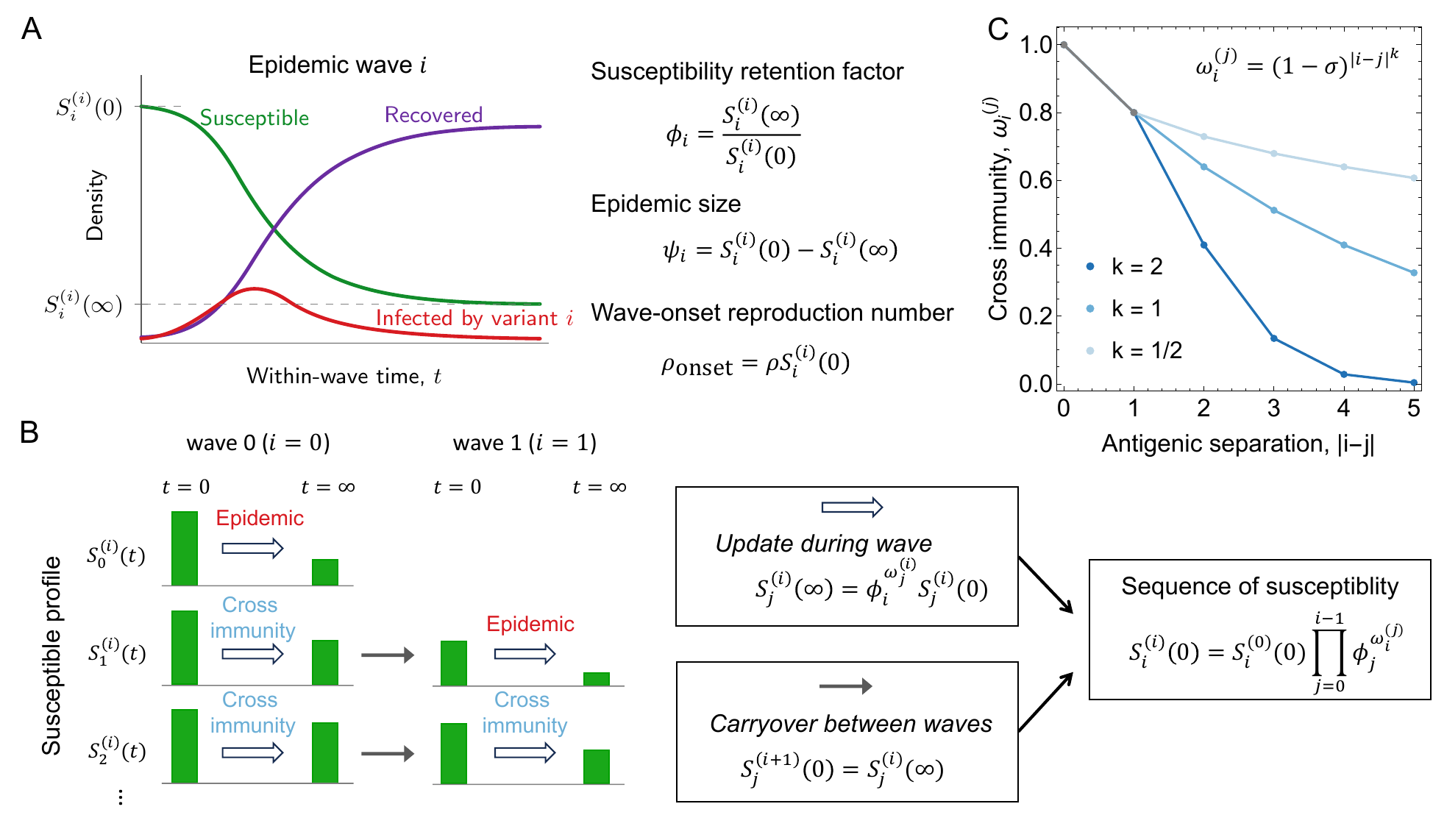}
    \caption{Schematic summary of our recurrent epidemic model. (A) During wave $i$, an SIR epidemic reduces the susceptible density to the current variant from $S_i^{(i)}(0)$ to $S_i^{(i)}(\infty)$. The susceptibility retention factor is $\phi_i=S_i^{(i)}(\infty)/S_i^{(i)}(0)$, and the epidemic size is $\psi_i=S_i^{(i)}(0)-S_i^{(i)}(\infty)=S_i^{(i)}(0)(1-\phi_i)$. (B) Each epidemic wave updates the susceptible profile for the current and future variants. During wave $i$, susceptibility to variant $j$ is reduced as $S_j^{(i)}(\infty)=\phi_i^{\omega_j^{(i)}}S_j^{(i)}(0)$, and this profile is carried over to the next wave. Iterating this update gives rise to the sequences of the susceptible density faced by each introduced variant. (C) Cross-immunity generated by variant $j$ against variant $i$ decreases with antigenic separation according to $\omega_i^{(j)}=(1-\sigma)^{|i-j|^k}$.}
    \label{fig:model}
\end{figure}

We first consider the changes in susceptibility to the epidemic variant (Fig.~\ref{fig:model}A). The densities of hosts susceptible to variant $i$ and currently infected by variant $i$ on day $t$ of wave $i$ are denoted by $S_i^{(i)}(t)$ and $I_i(t)$, respectively. The epidemiological dynamics within wave $i$ follow a simple SIR model,
\begin{align} \label{eq:SIRmodel}
    \frac{\mathrm{d}S_i^{(i)}(t)}{\mathrm{d}t}=-\beta S_i^{(i)}(t) I_i(t), \quad \frac{\mathrm{d}I_i(t)}{\mathrm{d}t}=\beta S_i^{(i)}(t) I_i(t)-\gamma I_i(t)
\end{align}
where $\beta$ and $\gamma$ are the infection and recovery rates, respectively. Recovered hosts acquire long-lasting immunity to the infecting variant. However, they may remain susceptible to antigenically distinct future variants, depending on the degree of cross-immunity. We assume that the interval between epidemic waves is long relative to the duration of a single wave, so that $I_i(\infty)=0$ before the next variant is introduced.

The epidemic wave changes the density of hosts susceptible to the current variant (Fig.~\ref{fig:model}A). The final density of hosts susceptible to variant $i$ in wave $i$ is
\begin{align}
    S_i^{(i)}(\infty)=\phi_i S_i^{(i)}(0),
\end{align}
where $\phi_i$ is the susceptibility retention factor, defined as the fraction of hosts who remain uninfected during wave $i$ among those susceptible at the beginning of the wave.
Here, $\phi_i$ can be defined by
\begin{align} \label{eq:phi}
    \phi_i&:=\begin{cases}
        \phi_i^* & (S_i^{(i)}(0)> S^*) \\
        1 & (S_i^{(i)}(0)\leq S^*), \\
    \end{cases}
\end{align}
where $S^{*}=1/\rho$ is the epidemic threshold, i.e., the minimum susceptible density required for an epidemic to grow, for basic reproductive number $\rho=\beta/\gamma$.
In addition, $\phi_i^*$ is given as the unique solution to
\begin{equation}
    \log\phi_i^*+\rho S_{i}^{(i)}(0)(1-\phi_i^*)=0
\end{equation}
that exists in the region $0<\phi_i^*<1$ \cite{andreasen2006shaping,ma2006generality}. Eq.~(\ref{eq:phi}) captures the threshold phenomenon: susceptibility is depleted when the initial susceptible density exceeds the epidemic threshold, whereas no epidemic occurs below the threshold~\cite{diekmann2013mathematical,kermack1927contribution,keeling2011modeling}.

Equivalently, this threshold condition can be expressed in terms of the wave-onset reproduction number,
\begin{equation}
    \rho_{\text{onset}}^{(i)}:=
    \rho S_i^{(i)}(0).
\end{equation}
This quantity measures the effective reproduction number faced by variant $i$ at the introduction to the population, after population immunity generated by previous waves has reduced the density of host susceptible to that variant. Thus, an epidemic occurs only when $\rho_{\text{onset}}^{(i)}>1$, whereas
$\rho_{\text{onset}}^{(i)} \leq 1$ corresponds to no epidemic.

We also define the final size of an epidemic in wave $i$, denoted by $\psi_i$, as the density of hosts that have ever been infected in the wave. $\psi_i$ is described as
\begin{align}
    \psi_i := \int_{0}^{\infty}\beta S_i^{(i)}(t) I_i(t)\,dt = S_i^{(i)}(0)-S_i^{(i)}(\infty) = S_i^{(i)}(0)(1-\phi_i).
\end{align}
Thus, $\psi_i$ is given by the product of the initial density of susceptible hosts, $S_{i}^{(i)}(0)$, and the proportion of the susceptible population that is ultimately infected in wave $i$, $1-\phi_i$.

\subsection{Protection against future variants due to cross-immunity}

We next evaluate how an epidemic wave generates immunity against closely related variants that appear later (Fig.~\ref{fig:model}B). Following a status-based polarized-immunity formulation in which cross-immunity is generated through the exposure to the infected hosts~\cite{gog2002dynamics,sasaki2022antigenic,kumata2022antigenic}, we assume that exposure to the hosts infected by variant $j$ induces complete protection against future variant $i$ with probability $\omega_i^{(j)}$ and leaves the host susceptible to variant $i$ with probability $1-\omega_i^{(j)}$. Let $S_i^{(j)}(t)$ denote the density of hosts susceptible to future variant $i\,(>j)$ at time $t$ during wave $j$. Under this formulation, the susceptible density changes according to
\begin{align}\label{eq:cross}
    \frac{\mathrm{d}S_i^{(j)}(t)}{\mathrm{d}t}=-\omega_i^{(j)} \beta S_i^{(j)}(t)I_j(t).
\end{align}
This equation should be interpreted as a status-based closure: rather than tracking the full joint distribution of immune histories, we track the marginal density of hosts susceptible to each future variant and assume that exposure to variant $j$ removes this susceptibility with probability $\omega_i^{(j)}$. Dividing Eq.~(\ref{eq:cross}) by Eq.~\eqref{eq:SIRmodel} and integrating over wave $j$ gives
\begin{align}
    S_i^{(j)}(\infty)=\phi_j^{\omega_i^{(j)}} S_i^{(j)}(0)
\end{align}
(see the Appendix for the detailed derivation). Thus one epidemic wave updates susceptibility not only to its own variant but also to each future variant against which it induces cross-immunity.

We also assume no waning of immunity between waves, so susceptibility at the beginning of a wave equals susceptibility at the end of the previous wave:
\begin{align}\label{eq:noloss}
    S_i^{(j+1)}(0)=S_i^{(j)}(\infty).
\end{align}
Combining Eqs.~(\ref{eq:cross}) and (\ref{eq:noloss}) gives the initial susceptible density in wave $i$ as a function of prior epidemics:
\begin{align}\label{eq:Sii}
    S_i^{(i)}(0)=S_i^{(0)}(0)\prod_{j=0}^{i-1}\phi_j^{\omega_i^{(j)}}
\end{align}
The host susceptibility to variant $i$ when wave $i$ starts is reduced to $\phi_j^{\omega_i^{(j)}}$-fold due to the past epidemic of wave $j$. The susceptibility to variant $i$ at the beginning of wave $i$, $S_i^{(i)}(0)$, is the product of the reduction by past waves from $j=0$ to $i-1$. The sequence $S_i^{(i)}(0)$ therefore summarizes the population immunity faced by each newly introduced variant in terms of the remaining susceptible density.

The above setup allows us to calculate recursively the sequence $S_i^{(i)}(0)$ from wave $0$. Given the sequence up to wave $i$, $\phi_0,\phi_1,\cdots,\phi_i$, one can calculate $S_{i+1}^{(i+1)}(0)$ from Eq.~(\ref{eq:Sii}) and then obtain $\phi_{i+1}$ by using Eq.~(\ref{eq:phi}). Repeating this process gives the sequences of susceptible densities $S_i^{(i)}(0)$, the within-wave susceptibility retention factors $\phi_i$, and the corresponding epidemic sizes $\psi_i$.

\subsubsection{Cross-immunity among variants}
Different epidemic variants have different but related antigenic properties. Therefore, immunity induced by one variant can also protect against another variant, but protection is weaker when the two variants are more antigenically different. As a simple approximation for gradual antigenic change, we use the order of introduction as a proxy for antigenic separation and assume that variants $j$ and $i$ are separated by $|i-j|$ antigenic steps. We assume that cross-immunity decreases with this separation as
\begin{align} \label{eq:omega}
    \omega_i^{(j)}=(1-\sigma)^{|i-j|^k},
\end{align}
where $\sigma$ measures the loss of cross-immunity between neighboring epidemic variants due to antigenic escape. The parameter $k$ controls how immune escape accumulates across antigenic steps. When $k=1$, cross-immunity decays multiplicatively with each step. When $k<1$, cross-immunity declines more slowly with antigenic separation. When $k>1$, it declines more rapidly.

\subsubsection{One-dimensional map}
When $k=1$, escape accumulates multiplicatively with antigenic distance. In this case, the effects of all earlier waves on variant $i+1$ can be collected into the susceptible density at the start of wave $i$, and the model reduces to the following one-dimensional map between successive waves (see the Appendix):
\begin{align}\label{eq:map}
    S_{i+1}^{(i+1)}(0)=(\phi_i S_{i}^{(i)}(0))^{1-\sigma}.
\end{align}
Equivalently, defining $x_i=S_i^{(i)}(0)$, we write $x_{i+1}=(x_i\phi_i)^{1-\sigma}$, where $\phi_i$ is determined by $\phi_i=\exp[-\rho x_i(1-\phi_i)]$ if $x_i>1/\rho$ and by $\phi_i=1$ otherwise. Thus $\sigma=0$ corresponds to complete cross-immunity between successive variants, whereas $\sigma=1$ corresponds to complete escape and independent waves. For intermediate $0<\sigma<1$, each wave is partly buffered by immunity generated in previous waves and partly renewed by antigenic escape. 

In the following analyses, we focus on how the partial cross-immunity shapes recurrent epidemic dynamics by numerically analyzing the full model in Eq.\eqref{eq:Sii}. We also conduct mathematical analysis of the one-dimensional map in Eq.~\eqref{eq:map} to support the numerical results.

\subsection{Numerical methods}
Numerical epidemic sequences were generated by iterating the recurrence for $S_i^{(i)}(0)$ rather than by explicitly integrating the within-wave ODEs. At each wave, $\phi_i$ was obtained from the final-size equation $\phi_i=\exp[-\rho S_i^{(i)}(0)(1-\phi_i)]$ when $\rho S_i^{(i)}(0)>1$, and was set to $\phi_i=1$ otherwise. The epidemic size was then calculated as $\psi_i=S_i^{(i)}(0)(1-\phi_i)$, and susceptibilities to future variants were updated according to Eq.~(\ref{eq:Sii}) with the cross-immunity kernel in Eq.~(\ref{eq:omega}). Long-term averages and bifurcation diagrams were computed after discarding transient waves; the period-classification procedure is described in the Appendix.

\section{Results}
\subsection{Wave-to-wave epidemic dynamics}
\begin{figure}
    \centering
    \includegraphics[width=1\linewidth]{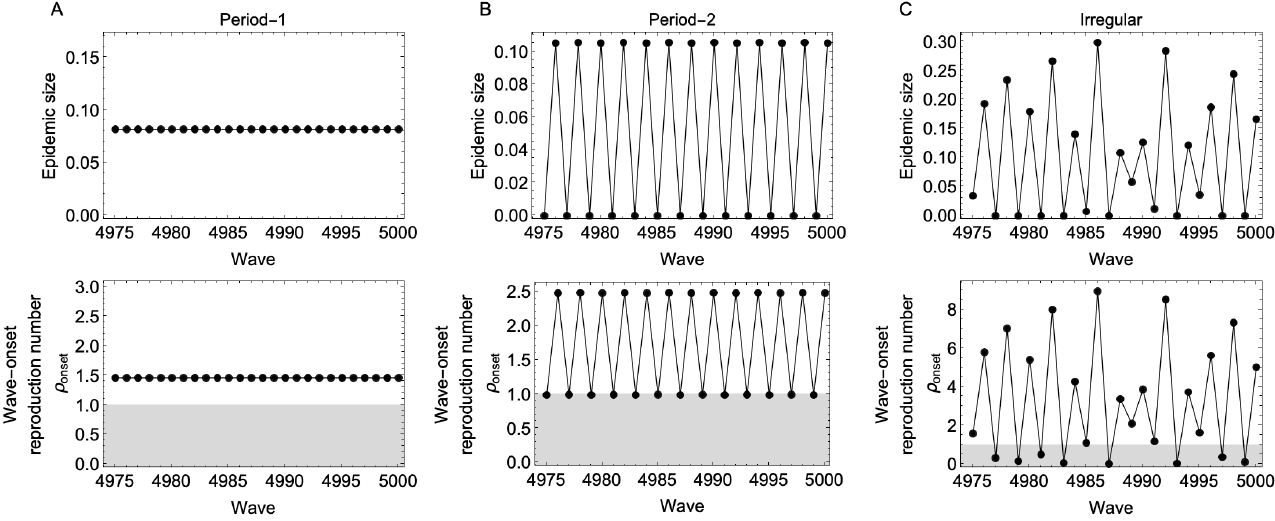}
    \caption{Representative recurrent epidemic dynamics. Upper panels show epidemic size, $\psi_i$, and lower panels show the wave-onset reproduction number, $\rho_{\mathrm{onset}}^{(i)}=\rho S_i^{(i)}(0)$, across successive introduced variants. (A) Period-$1$ dynamics, with equal-sized recurrent epidemics ($k=1$, $\rho=10$). (B) Period-$2$ dynamics, with alternating large and negligible epidemics ($k=1$, $\rho=21$). (C) Irregular dynamics consistent with high-period or chaos-like behavior ($k=2$, $\rho=30$). The shaded region indicates $\rho_{\mathrm{onset}}^{(i)}\leq 1$, where no epidemic occurs. Other parameter: $\sigma=0.3$.}
    \label{fig:seq}
\end{figure}

To illustrate how immune history generates distinct wave-to-wave regimes, we first show representative trajectories (Fig.~\ref{fig:seq}). For $k=1$ and moderate $\rho$, both epidemic size, $\psi_i$, and the wave-onset reproduction number, $\rho_{\mathrm{onset}}^{(i)}$, converge to constant values, yielding period-$1$ recurrence with equal-sized epidemics (Fig.~\ref{fig:seq}A). At larger $\rho$, this fixed point loses stability: a large epidemic is followed by a wave for which $\rho_{\mathrm{onset}}^{(i)}$ falls to the epidemic threshold, resulting in a period-$2$ sequence of large and negligible epidemics (Fig.~\ref{fig:seq}B). When cross-immunity declines rapidly with antigenic distance ($k=2$), the sequence can instead become irregular, with epidemic sizes and wave-onset reproduction numbers varying widely across successive waves (Fig.~\ref{fig:seq}C). These regimes arise because a large epidemic strengthens cross-immunity against nearby future variants, whereas antigenic escape progressively restores susceptibility to later variants.

\subsection{Bifurcation diagrams}
\begin{figure}
    \centering
    \includegraphics[width=1\linewidth]{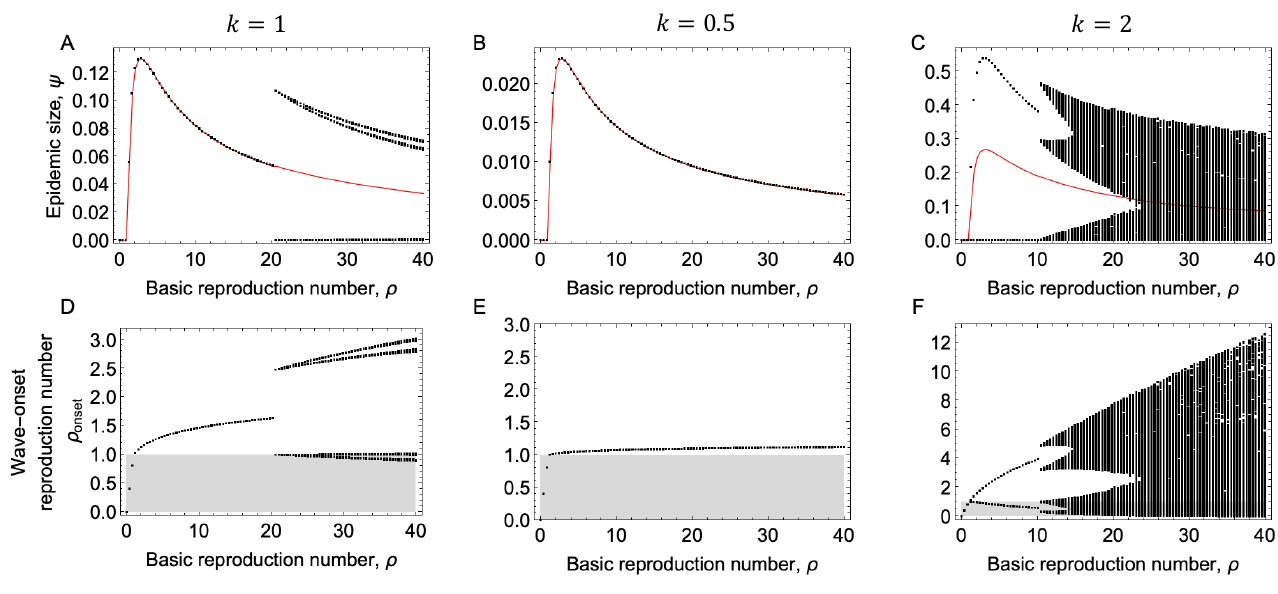}
    \caption{Bifurcation diagrams of recurrent epidemic size (A, B, C) and wave-onset reproduction number (D, E, F). Black points show long-term epidemic sizes sampled over 500 points after transients for each $\rho$, and red curves show the mean epidemic size. Light gray region indicates no-epidemic conditions ($\rho S_i^{(i)}(0)<1$). Panels compare different values of the cross-immunity shape parameter: (A, D) $k=1$, (B, E) $k=0.5$, and (C, F) $k=2$. Other parameters: $\sigma=0.3$}
    \label{fig:bif}
\end{figure}

Next, we examined the long-term behavior of this sequence. In the bifurcation diagrams of epidemic size in Fig.~\ref{fig:bif}A-C, black points show the values of the epidemic size $\psi_i$ for each value of $\rho$, whereas red curves show the mean over the same sequence. 

Fig.~\ref{fig:bif}A shows the bifurcation diagram when $k=1$ and $\sigma=0.3$. No epidemic occurs for $\rho<1$, where $\phi_i = 1$ with our assumption. For $1<\rho\lesssim 20$, epidemics occur and their sizes converge to a single value, corresponding to period-$1$ recurrent epidemics of equal size. When $\rho$ exceeds this range, the period-$1$ solution loses its stability, and a period-$2$ cycle emerges (period-doubling bifurcation). At $\rho\approx 21$, the black points split into nonzero and near-zero branches. The near-zero point corresponds to waves whose epidemic size is effectively reduced by population immunity left by earlier epidemics. In this case, for larger $\rho$, a further period-doubling bifurcation yields clear period-$4$ patterns (e.g. $\rho=30$). Thus, strong feedback through cross-immunity can destabilize equal-sized recurrent epidemics when the basic reproduction number is large, giving rise to recurrent waves of unequal size.

The shape of cross-immunity strongly affects the bifurcation pattern. When $k=0.5$, cross-immunity from earlier variants tends to remain effective against later variants with greater antigenic separations, and population immunity generated by past epidemic waves therefore has a long-lasting effect on future waves. In this case, epidemic sizes converge to a single recurrent value even for large values of $\rho$ (Fig.~\ref{fig:bif}B). In contrast, when $k=2$, cross-immunity from earlier variants declines rapidly for later variants. Population immunity generated by past waves then has a weaker effect on future waves. The recurrent epidemics with period-$2$ is observed at much smaller $\rho$ ($\simeq 1$) (Fig.~\ref{fig:bif}C). Around $\rho=10$, the period-$2$ cycle loses stability and is replaced by higher-period dynamics. A representative trajectory at $\rho=12$ shows that the resulting sequences can contain recurring large, intermediate, and small waves with an additional longer-period modulation (Fig.~S1). At $\rho=25$, the dynamics exhibit broadly distributed epidemic sizes with chaos-like behavior (Fig.~S1). Despite these marked differences in the bifurcation pattern, the mean epidemic size (red curves) shows a similar trend across panels with peaks at the intermediate $\rho$.

Furthermore, population immunity strongly constrains the wave-onset reproduction number,
$\rho_{\mathrm{onset}}^{(i)}$, which is the effective reproduction number
experienced by a variant at introduction, rather than the intrinsic basic
reproduction number $\rho$ (Fig.~\ref{fig:bif}D-F). For example, at $k=1$, the bifurcation point of period-$1$ corresponds to a $\rho$ of approximately $20$, and this indeed corresponds to a wave-onset reproduction number of around $1.5$ (Fig.~\ref{fig:bif}D). Thus, even when the intrinsic basic reproduction number is large, the effective reproduction number at wave onset remains close to the epidemic threshold because population immunity strongly reduces susceptibility.
These above comparisons of bifurcation diagrams reveal that patterns of recurrent epidemic dynamics depend on the shape of the cross-immunity kernel $k$.

\subsection{Periodicity of recurrent epidemics}
To characterize periodicity across parameter space, we numerically simulated epidemic sequences and classified their dominant period using a Fourier-based procedure (see the Appendix). Figure~\ref{fig:period} summarizes the observed periods across $\rho$ and $\sigma$. The black region at low $\rho$ corresponds to parameter combinations for which the epidemic cannot persist, whereas the light-blue region corresponds to period-$1$ recurrence. For $k=0.5$, period-$1$ recurrence occupies almost the entire parameter space, except for a small region at weak escape and moderate to high $\rho$ (Fig.~\ref{fig:period}A). For $k=1$, the stable period-$1$ region gives way to period-$2$ and period-$4$ dynamics when $\rho$ is large and antigenic escape is sufficiently weak (Fig.~\ref{fig:period}B). For $k=2$, these higher-period regimes expand substantially and include broad regions classified as period-$2$, period-$4$, and higher periods, including complex recurrent dynamics (Fig.~\ref{fig:period}C). Therefore, the periodicity of recurrent epidemics depends strongly on the shape of the cross-immunity kernel.

For $k=1$, the one-dimensional map, Eq.~\eqref{eq:map}, allows us to investigate the analytical stability for periodic solutions (see the Appendix). Let $f(x)$ be the return map for the initial susceptible density. The stability of a fixed point is determined by whether the absolute value of the derivative, $|\mathrm{d}f/\mathrm{d}x|$, is smaller or larger than one. The period-$1$ recurrent epidemic loses stability when
\begin{equation}
    \left|\frac{\mathrm{d}f}{\mathrm{d}x}(x^*)\right|=1
\end{equation}
where $x^*$ is the fixed point. The orange dashed curve in Fig.~\ref{fig:period}B shows this analytical boundary and agrees well with the numerical classification. In the limit $\sigma\to0$, the boundary intersects the $\rho$ axis at $\rho=e^3$, indicating that destabilization of period-$1$ recurrence requires very large $\rho$ when antigenic escape is weak (see the Appendix). For the second iterate $g(x)=f\circ f(x)$, the stability of a period-$2$ orbit is determined by the derivative of $g$ evaluated at the points on that orbit. For the doubly iterate map $g(x)=f\circ f(x)$, a period-$2$ cycle loses stability when
\begin{equation}
    \left|\frac{\mathrm{d}g}{\mathrm{d}x}(x^*)\right|=1
\end{equation}
at the fixed point of $g(x)$. The red dashed curve calculated from the above condition captures well the bifurcation region for large $\sigma$ (upperside of the boundary between period-$2$ and period-$4$) (Fig.~\ref{fig:period}B). 
However, for small $\sigma$, the analytical and numerical boundaries differ
slightly. In this region, the period-$4$ orbit consists of two nearly
indistinguishable period-$2$-like subcycles (Fig.~S2). Because the difference
between these subcycles is very small, identifying the trajectory as
period-$4$ requires a stringent numerical tolerance. Thus, the apparent
discrepancy primarily reflects the finite resolution of the numerical
period-classification procedure. Altogether, the analytical and numerical results show that recurrent epidemic dynamics can undergo a cascade from period-$1$ recurrence to higher-period or complex recurrent behavior.

\begin{figure}
    \centering
    \includegraphics[width=1\linewidth]{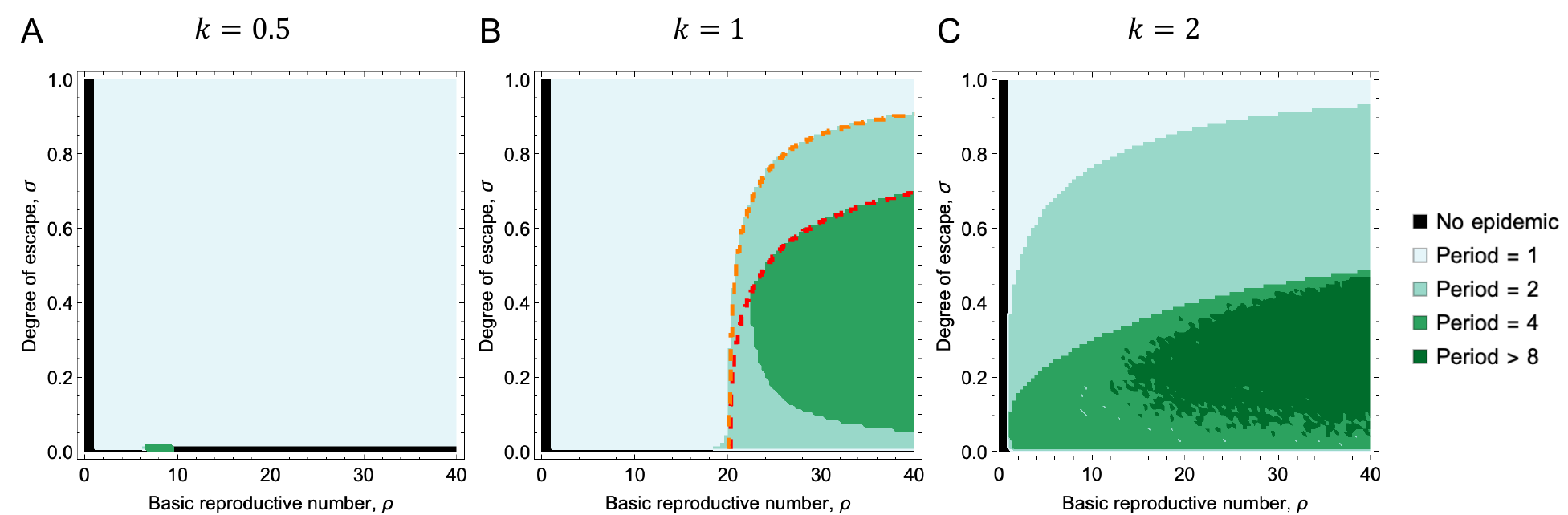}
    \caption{Periodicity of recurrent epidemic dynamics across $\rho$ and $\sigma$. Colors indicate the dominant period estimated from simulated epidemic sequences. Black indicates no persistent epidemic. Dashed curves in panel B show analytical stability boundaries for the period-$1$ and period-$2$ solutions. Numerical methods for periodicity detection are explained in the Appendix.}
    \label{fig:period}
\end{figure}

\subsection{Maximization of epidemic size at an intermediate $\rho$}
\begin{figure}
    \centering
    \includegraphics[width=1\linewidth]{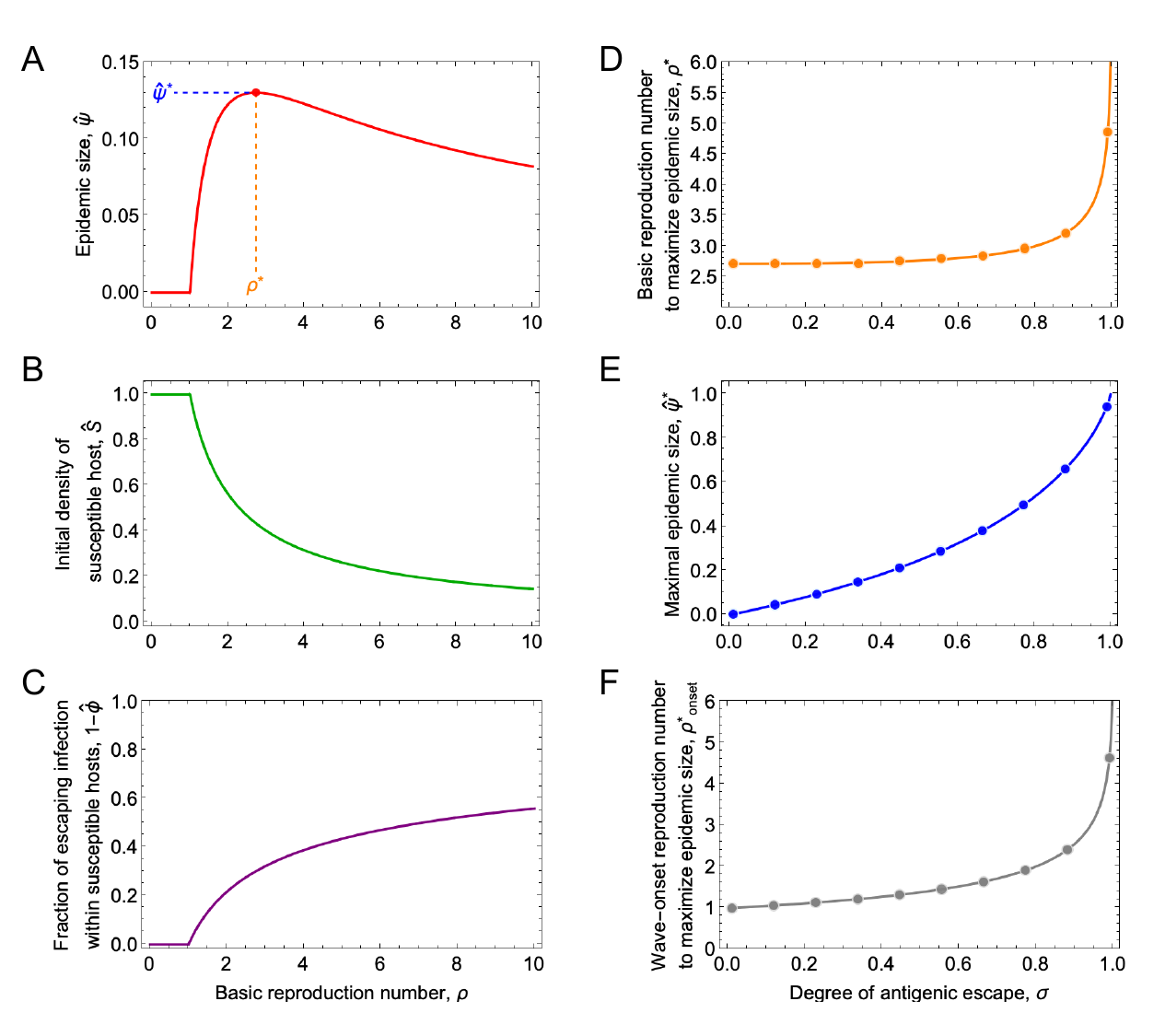}
    \caption{Intermediate basic reproduction numbers maximize recurrent epidemic size for $k=1$. (A) recurrent epidemic size $\hat{\psi}$ as a function of $\rho$, (B) initial susceptible density $\hat{S}$, and (C) infection probability among susceptible hosts, $1-\hat{\phi}$. Dashed lines in panel A indicate $\rho^*$ and $\hat{\psi}^*$. Panels D and E show how (D) the maximizing basic reproduction number $\rho^*$ and (E) the maximal recurrent epidemic size $\hat{\psi}^*$ vary with $\sigma$. (F) Wave onset reproduction number $\rho^*_{\text{onset}}$ that is realized value of $\rho_{\text{onset}}$ when $\rho$ takes the value $\rho^*$ that maximizes recurrent epidemic size for a given $\sigma$. The lines are analytical solutions and points are obtained from the numerical simulation of the dynamics. Parameter: $\sigma=0.3$}
    \label{fig:episize}
\end{figure}

The mean epidemic size reaches a maximum at intermediate $\rho$ and then decreases as $\rho$ increases (red lines in Fig.~\ref{fig:bif}). In the numerical simulations, this non-monotonic pattern was observed across the three values of $k$ examined. For analytical clarity, we focus below on the stable period-$1$ regime for $k=1$. 
A central result is that the recurrent epidemic size is maximized at an intermediate basic reproduction number (Fig.~\ref{fig:episize}A). This contrasts with the final-size relation for a single SIR epidemic, where larger $\rho$ generally increases outbreak size. For the representative case $\sigma=0.3$ shown in Fig.~\ref{fig:episize}A-C, $\hat{\psi}$ first rises after epidemics become possible, reaches its maximum at $\rho^*$, and then declines even though transmission within each wave becomes stronger.

The mechanism follows from the decomposition
\begin{align} \label{eq:psi}
    \hat{\psi} = \hat{S}(1-\hat{\phi}),
\end{align}
where hats denote the recurrent equilibrium. The factor $1-\hat{\phi}$ is the infection probability among hosts susceptible at the start of a wave. It increases with $\rho$ because, for a given initial susceptible density, a larger basic reproduction number produces a larger epidemic size in a single wave (Fig.~\ref{fig:episize}C). In contrast, $\hat{S}$ is the susceptible density inherited by the newly introduced variant. It decreases with $\rho$ because larger preceding epidemics leave stronger cross-immunity against later related variants (Fig.~\ref{fig:episize}B). Thus increasing $\rho$ has two opposing effects: it makes infection more efficient within the current wave, but it also depletes the susceptible density available to future waves. Near the invasion threshold ($\rho \approx 1$), many hosts remain susceptible, but each susceptible host has a low infection probability. Conversely, at large $\rho$, susceptible hosts are infected efficiently within a wave, but the susceptible density available to the next antigenically related variant is strongly depleted. It therefore reaches a maximum at an intermediate $\rho$ (Fig.~\ref{fig:episize}A).

We next ask how the maximal epidemic size shifts with the degree of antigenic escape. We analytically calculated the basic reproduction number $\rho^*$ that maximizes $\hat{\psi}$, and the corresponding maximal epidemic size $\hat{\psi}^*$, as functions of $\sigma$ (see the Appendix):
\begin{equation}
    \rho^*=(1-\sigma)^{1-\frac{1}{\sigma}} \cdot \frac{-\log(1-\sigma)}{\sigma}
\end{equation}
and
\begin{equation}
    \hat{\psi}^*=\frac{\sigma}{(1-\sigma)^{1-\frac{1}{\sigma}}}.
\end{equation}
At this maximal value, the wave-onset reproduction number also takes the simple form
\begin{equation}
    \rho^*_{\text{onset}}=\frac{-\log(1-\sigma)}{\sigma}.
\end{equation}

These analytical results match the numerical values shown in Fig.~\ref{fig:episize}D-F. When antigenic escape is weak, even a relatively small value of $\rho$ can generate enough cross-immunity to reduce the susceptible density available to later variants. In the weak escape limit,
\begin{equation}
    \lim_{\sigma \to 0} \rho^* = e
\end{equation}
and the expansion $\rho^*=e+(e/24)\sigma^2+O(\sigma^3)$ shows that the first-order effect of $\sigma$ vanishes (see the Appendix). Thus, $\rho^*$ remains close to $e$ when antigenic escape is small (Fig.~\ref{fig:episize}D). When $\sigma$ is large enough, immunity generated by earlier variants protects the hosts less effectively against infection by later variants. The $\rho$ that maximizes recurrent epidemic size therefore shifts upward and diverges as $\sigma\to1$. At the same time, the maximal recurrent epidemic size increases monotonically with $\sigma$ because stronger antigenic escape leaves a larger susceptible pool for each newly introduced variant, allowing larger recurrent epidemics (Fig.~\ref{fig:episize}E). The wave-onset reproduction number at which recurrent epidemic size is maximized also increases with $\sigma$, but remains only slightly above the epidemic threshold for small or intermediate antigenic escape (Fig.~\ref{fig:episize}F).

\section{Discussion}
The dynamic interplay between population immunity and antigenic escape of pathogens is a key factor shaping epidemiological dynamics. Here, we developed a recurrent epidemic model in which immune history links successive epidemic waves of antigenically related variants. The model shows that equal-sized recurrent epidemics are stable over broad parameter ranges, but can be destabilized into period-$2$, period-$4$, or more complex dynamics when transmission is strong and antigenic escape is limited. It also reveals a non-monotonic relationship between the basic reproduction number and recurrent epidemic size: recurrent epidemic size is maximized at an intermediate value of $\rho$.

For the analytically tractable case $k=1$, this maximum arises from negative inter-wave feedback through cross-immunity. Increasing $\rho$ raises the infection probability within the current wave, but also strengthens immunity against antigenically related future variants, reducing the susceptible density available to later waves. This distinction helps interpret reproduction numbers estimated from empirical influenza data. Because reproduction numbers estimated from epidemic data depend on population context and immunity \cite{delamater2019complexity}, such estimates in populations with pre-existing immunity may be more closely related to the wave-onset quantity $\rho_{\text{onset}}$ than to the intrinsic $\rho$. Reported influenza reproduction numbers are typically modest, with median estimates of $1.28$ for seasonal influenza and $1.46$ for the 2009 pandemic, and most reported values lie between $1$ and $2$ \cite{biggerstaff2014estimates}. The value of $\rho^*_{\text{onset}}$, at which the averaged epidemic size is maximized, lies in this range for a wide range of (small to intermediate) antigenic escape (Fig.~\ref{fig:episize}F). This suggests that recurrent epidemic size can be maximized in a parameter range where the wave-onset reproduction number remains comparable to values commonly estimated from influenza epidemic data.

A related non-monotonic effect of transmissibility on epidemic size has been reported in the SIRC model, where higher contact rates can increase immune boosting and thereby prolong cross-protection \cite{casagrandi2006sirc}. In our model, the mechanism is different: increasing $\rho$ raises the infection probability within the current wave, but the cross-immunity generated by that wave reduces susceptibility to later antigenically related variants.

The shape of cross-immunity further influences the periodic behavior of recurrent epidemics. When immune escape accumulates slowly with antigenic distance, past waves continue to buffer later variants, and equal-sized recurrence tends to remain stable. When escape accumulates rapidly, the immune footprint of earlier waves becomes
shorter-ranged in antigenic space, and multi-periodic dynamics arise more easily. Thus, quantitative details of cross-immunity and antigenic maps are important for understanding and predicting the epidemiological dynamics \cite{smith2004mapping,fonville2014antibody,katzelnick2021antigenic,markov2023evolution,ito2025integrative}.

Our model is related to season-to-season and multi-strain models of influenza dynamics, which have examined how epidemic size, cross-immunity, antigenic drift, and host demography jointly shape recurrent outbreaks \cite{andreasen2003dynamics,boni2004influenza,roberts2019simple,andreasen2006shaping}. In particular, Boni et al.\ \cite{boni2004influenza} considered a positive feedback between epidemic size and antigenic escape: larger epidemics generate greater antigenic change, which in turn allows later variants to evade accumulated immunity and cause larger epidemics. Our model instead isolates a distinct negative inter-wave feedback mediated by cross-immunity. Conditioned on a given sequence of antigenically related variants, a large epidemic leaves stronger cross-immunity and thereby reduces the susceptible density available to subsequent variants. Thus, rather than asking how epidemic size drives antigenic escape, we ask how immune history constrains the sizes of later epidemics once antigenic relationships among variants are specified.

This focus also defines the main limitations of the model. We prescribe the order and antigenic separation of introduced variants, rather than modelling their mutation, migration, establishment, or competition explicitly. Consequently, the model does not predict which variant will arise next, how rapidly antigenic change occurs, or how the positive feedback between epidemic size and antigenic escape may interact with the negative immune feedback identified here. We further assume that epidemic waves are temporally separated and that immunity does not wane between waves. These assumptions allow the wave-to-wave effect of cumulative cross-immunity to be isolated, but exclude overlapping circulation, immune waning, vaccination, and demographic turnover.

A further important assumption in our framework is the status-based representation of immunity. The model tracks the marginal density of hosts susceptible to each future variant, rather than the full joint distribution of immune histories across variants \cite{kryazhimskiy2007state,wikramaratna2015five}. In this formulation, exposure to a variant confers complete protection against a future variant for a fraction of hosts, while leaving the remaining hosts fully susceptible. Alternative immunological assumptions, such as leaky cross-immunity that partially reduces susceptibility in all previously exposed hosts, could generate different patterns of cumulative protection and thereby alter the stability and periodicity of recurrent epidemics \cite{reyne2025leaky}. Correlations among susceptibilities to different variants are also compressed into the cross-immunity update, which may affect the quantitative accumulation of protection across antigenic distances. An important next step will be to examine the robustness of the feedback identified here across alternative immune mechanisms, while incorporating stochastic antigenic emergence, explicit immune histories, and continuous-time co-circulation.

Another important direction is to move beyond a single focal population. In a metapopulation setting, different host populations could experience different immune histories and exchange variants through migration, so that local immune history and global antigenic circulation jointly shape recurrent epidemic dynamics~\cite{russell2008global,viboud2006synchrony,bedford2010global}. Finally, allowing transmissibility and antigenic escape to evolve jointly could help clarify whether pathogen evolution is driven primarily by short-term epidemic advantage or by longer-term recurrent epidemic success \cite{sasaki2022antigenic}.

Overall, recurrent epidemics cannot be understood by repeatedly applying single-epidemic intuition. When successive variants are linked by cross-immunity, the immune footprint left by one epidemic becomes part of the epidemiological environment encountered by later variants. This inter-wave feedback can stabilize equal-sized recurrence, destabilize it into multi-periodic or complex dynamics, and make mean recurrent epidemic size largest at intermediate transmissibility.

\bibliographystyle{cas-model2-names}
\bibliography{biblio.bib}

\end{document}